\documentclass[twocolumn,prl,showpacs]{revtex4}  
\usepackage{amssymb}

\usepackage{epsfig}

\begin{document}

\title{ Anomalous diffusion in supercooled liquids: a long-range
  localisation in particle trajectories}

\author{T. Oppelstrup$^1$ and M. Dzugutov$^2$}

\affiliation{$^1$ Dept. for Numerical Analysis and Computer
Science\\ $^2$ Dept. for Materials Science and Engineering\\
Royal Institute of Technology, S--100 44 Stockholm, Sweden}
\begin{abstract}
  A statistical analysis of the geometries of particle trajectories in
  the supercooled liquid state is reported. We examine two structurally
  different fragile glass-forming liquids simulated by molecular
  dynamics. In both liquids, the trajectories are found to exhibit a
  long-range localisation distinct from the short-range localisation
  within the cage of nearest neighbours.  This novel diffusion anomaly
  is interpreted as a result of the potential-energy landscape
  topography of fragile glass-formers where the local energy minima
  coalesce into metabasins - compact domains with low escape
  probability.
\end{abstract}

\date{\today}


\maketitle

It is generally understood that the non-Arrhenius behaviour and other
dynamical anomalies observed in fragile \cite{Angell91} glass-forming
liquids are caused by a particular topography of the potential
energy-landscape (PEL) \cite{Debenedetti, Goldstein}. Having been
cooled below the cross-over temperature $T_A$, the liquid stays within
the basins of attraction of local PEL minima long enough for the
velocities to decorrelate, and its structural relaxation is controlled
by the inter-basin connectivity network. A distinctive feature of the
PEL of a fragile liquid is that the basins are geometrically organised
into a two-scale pattern whereby compact sets of contiguous basins
(metabasins)\cite{Stillinger, Heuer} are separated by barriers with
higher activation energy than those for elementary interbasin
transitions.

The impact of this PEL pattern upon the supercooled liquid dynamics
can be understood as follows. In a multidimensional space, a boundary
is close to almost every point of the domain it confines
\cite{Ma}. Therefore, almost every PEL minimum in a metabasin
is separated by the metabasin's boundary from some of its Euclidean
neighbour minima \cite{Dzugutov, Dzugutov2}. The resulting reduction
in the number of {\it accessible} relaxational degrees of freedom
slows down the relaxation and renders elementary interbasin
transitions increasingly collective \cite{Donati, Buldyrev}, in
accordance with the Adam-Gibbs conjecture \cite{Adam}. At a larger
time scale, the system's confinement to a metabasin is expected to
produce a localisation effect in its configuration-space
trajectory. It is assumed that the intra-metabasin dynamics correspond
to $\beta$-relaxation, whereas irreversible $\alpha$-relaxation, and
the respective decorrelation of the system's configuration-space
trajectory, occur by virtue of the inter-metabasin
transitions \cite{Stillinger}. In contrast to collective
intra-metabasin dynamics, relaxation beyond the metabasin boundaries
is concluded to be mediated by uncorrelated movements of particles
\cite{Kob}.

The configuration-space trajectory of a liquid can be investigated by
means of statistical analysis of its 3D-space projection - a
trajectory of a single particle. Based on the above arguments, we can
discriminate three distinct regimes in the particle trajectory of a
supercooled liquid (following the initial ballistic regime). (i)
Short-range localisation of a particle within the cage of its
neighbours; (ii) Persistent diffusion due to the collective
intra-metabasin dynamics \cite{Douglas}; (iii) Long-range
localisation produced by the the system's long-time confinement to a
metabasin. These anomalies can be detected as the respective
deviations of Hausdorff measure from its value for the Brownian
diffusion \cite{Mandelbrot}.  Regimes (i) and (ii) have been
observed in molecular-dynamics simulations \cite{Douglas}; the third
one has not been reported so far.

In this Letter, we analyse the statistical geometry of particle
trajectories in the supercooled liquid state. We examine two structurally
distinct fragile glass-forming liquids simulated by molecular
dynamics. Evidence is presented for the long range localisation in the
particles diffusion conjectured above as regime (iii). The observed
effect can be regarded as direct evidence for the presence of
metabasins in the PEL topography of fragile liquids. It also makes it
possible to assess the spatial extent of a metabasin. We discuss the
impact of the metabasin confinement on the relation between relaxation
and diffusion.

The MD simulations we report here explore two simple fragile
glass-formers. One is the Z2 model \cite{Z2} demonstrating a
pronounced tendency for icosahedral clustering. This liquid was
simulated at the number density $\rho=0.85$. At that density, its
estimated mode-coupling theory \cite{Gotze} critical temperature
$T_c=0.65$, fragility index $B=4.5$ and $T_A=1.1$. The liquid was
simulated at two temperatures: $T=1.2$ and $T=0.7$, above and below
$T_A$, respectively. The other glass-forming liquid explored here was
the binary Lennard-Jones system (BLJ) commonly used in glass studies
\cite{kob-andersen-blj, wales, glotzer-blj}. The interaction potential
was truncated at $r=3.0036$ using the Stroddard-Ford quadratic cutoff
\cite{ford}. The BLJ liquid was simulated in a supercooled state at
$\rho=1.19$ and $T=0.451$. In this model, we analysed the trajectories
of the smaller (more mobile) particles. Both models were simulated in
the $NVE$ ensemble with $N=16000$.

Traditionally, statistics of diffusing particles is presented in
terms of the distribution of particle displacements at time $t$,
$G_s(r,t)$ \cite{Hansen}. In the case of Brownian diffusion, this
is a Gaussian, and its second moment, the mean-square
displacement, depends on $t$ as $\langle r(t)^2\rangle = 6Dt$,
$D$ being the diffusion coefficient.

Alternatively, the particles' diffusion can be described in terms
of the first-passage time distribution $P(\varepsilon,t)$
\cite{Lindenberg}.  $P(\varepsilon,t) dt$ is the probability that
a particle's first crossing of the boundary of a sphere of
radius $\varepsilon$ centered at the particle's initial position
happens within the time interval between $t$ and $t+dt$. For
Brownian diffusion, the first moment of this distribution, the
mean first-passage time $ \tau_\varepsilon $ is equal to
$\tau_\varepsilon = \varepsilon^2/ (6D)$ \cite{Lindenberg}.

In this study, we analyse the statistical geometry of particle
trajectories. It is convenient for this purpose to use the
distribution of first passage trajectory lengths (FPL)
$P(\varepsilon,L)$, rather than first passage times. These two
distributions are equivalent for the description of liquid diffusion
(except for very short times) since, in dense liquids, velocities
decorrelate much faster than trajectories. The length of a particle
trajectory is defined as the time integral of its instantaneous
speed. With good accuracy, the trajectory length within time $t$ is
$L=v t$, where $v=\sqrt{8T/\pi}$ is the mean particle speed
\cite{Mayer}. We denote the first moment of the FPL as
$L_\varepsilon$. For the trajectory of a Brownian particle we have
\begin{equation}\label{Leps2} 
 L_\varepsilon  =  \frac{v\varepsilon^2}{6D}
\end{equation}

In this study we explore correlations in the system's
configuration-space trajectory and use the decay of the
correlations as an indicator of the system's approach to ergodic
equilibrium. The presence of correlation effects in a particle
trajectory can be detected from its scaling behaviour. The latter
can be quantitatively analyzed in terms of the trajectory's
fractal dimensionality (Hausdorff measure) \cite{Mandelbrot}. The
Hausdorff measure $\alpha$ is defined as the logarithmic
derivative of $L_\varepsilon$:
\begin{equation}\label{haus}
\alpha(\varepsilon)  =  \frac{d \ln L_\varepsilon}{d \ln \varepsilon}
\end{equation}

As it follows from equations (\ref{Leps2}) and (\ref{haus}),
$\alpha(\varepsilon)=2$ for a particle trajectory in the
large-$\varepsilon$ limit of Brownian diffusion.

Within the intermediate range of $\varepsilon$, the deviation of
$\alpha(\varepsilon)$ from the indicated Brownian value can be
used to discriminate the specific diffusion regimes discussed
above. If $\alpha(\varepsilon) > 2$, consecutive particle
displacements, defined by the scale of $\varepsilon$, correlate
negatively (the diffusion is localised).  If $\alpha(\varepsilon)
< 2$, consecutive particle displacements correlate positively;
this is a signature of persistent diffusion.

Fig. 1 shows $\alpha(\varepsilon)$ for the three liquid states
simulated in this study. A common feature in all cases is a
pronounced peak located at around $\varepsilon = 0.3$ where
$\alpha(\varepsilon)$ exceeds the Brownian value
$\alpha(\varepsilon)=2$. The peak, greatly enhanced by cooling
below $T_A$, indicates negative correlations in the particle
trajectory within the respective range of $\varepsilon$. It
apparently represents a localisation effect induced by the
particle's confinement to the cage of nearest neighbours. This
can be seen as a 3D projection of the configuration-space
trajectory localisation within a PEL minimum basin.

For the Z2 liquid at $T=1.2$, representing normal liquid
dynamics, the described short-range localisation regime is
immediately followed at larger $\varepsilon$ by convergence to
the asymptotic Brownian behaviour. In the case of the supercooled
liquid dynamics, both for the Z2 liquid at $T=0.7$ and the BLJ
liquid, a broad minimum is observed within $0.8<\varepsilon<3$
where $\alpha(\varepsilon)<2$. This effect, indicating positive
correlations in a particle trajectory, is characteristic of
persistent diffusion (regime (ii) discussed above). A similar
behaviour has already been reported for the BLJ liquid, and
concluded to be related to the collective particle motions
\cite{Douglas}.

At still larger values of $\varepsilon$ we observe, both for the
Z2 liquid at $T=0.7$ and the BLJ liquid, another anomaly in the
behaviour of $\alpha(\varepsilon)$. Before eventually converging
to the Brownian limit, $\alpha(\varepsilon)$ in these supercooled
liquids exhibits a second maximum located within
$3<\varepsilon<10$ where $\alpha(\varepsilon)>2$. This feature,
shown in the inset in Fig. 1, indicates the presence of a
long-range localisation in the particle diffusion that we
conjectured above as regime (iii). This interesting novel effect
is the central observation of the present study, and we shall
discuss it in more detail.

Besides Brownian diffusion, another essential criterion of the
ergodic behaviour of a liquid is homogeneity of the diffusion
process. In terms of the statistics of the particles'
trajectories, this can be measured by comparing the variance of
the FPL distribution: $Var[P(L,\varepsilon)] = \langle
L(\varepsilon)^2\rangle/L_\varepsilon^2 - 1$ with its asymptotic
value of $2/5$ expected for Brownian diffusion
\cite{Lindenberg}. The evolution of this quantity with respect to
$\varepsilon$ for all three simulated liquid states is shown in
Fig. 2. Comparing these data with those in Fig. 1 makes it
possible to conclude that the anomalies in the fractal
dimensionality observed in the supercooled liquid state, both for
Z2 and BLJ, are accompanied by anomalously high levels of
variance of the FPL distribution. Excessive variance is observed
in both the localisation and the persistent diffusion
domains. This observation is consistent with the point of view
that the anomalous diffusion in the supercooled liquid state is
associated with dynamical heterogeneity.

As we argued above, the localisation regime that we observed as a
large-$\varepsilon$ maximum in $\alpha(\varepsilon)$ for the
supercooled liquid state in Fig. 1 can be viewed as a result of the
system's confinement to a configuration-space domain that we associate
with a metabasin. In this way, the effect in question is a direct
observation of the constraints imposed by the metabasin structure of
the PEL upon the liquid diffusion, and its location in terms of
$\varepsilon$ makes it possible to conclude the extent of that
structure.

\begin{figure} %figure 1
\includegraphics[width=7.9cm]{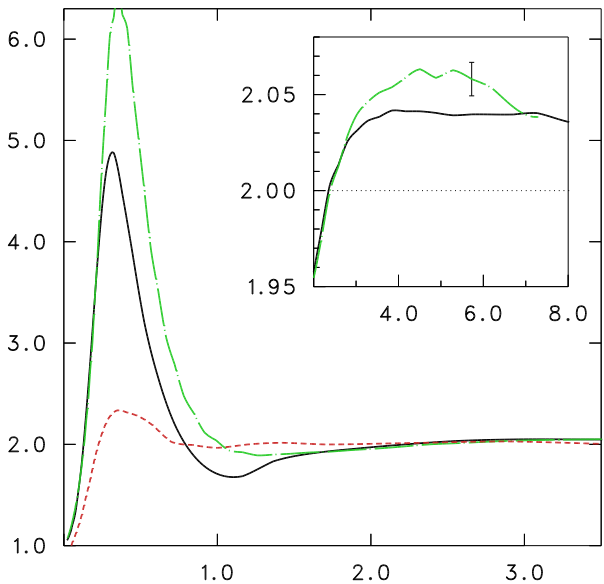}
\caption{ Apparent fractal dimension of a particle
  trajectory. Solid black line and dashed red line: Z2 liquid for
  $T=0.7$ and $T=1.2$, respectively. Dash-dotted green line, the
  BLJ liquid. The error bar indicate the estimated standard
  deviation for the BLJ data.}
\label{fig1}
\end{figure}

\begin{figure} %figure 2
\includegraphics[width=7.cm]{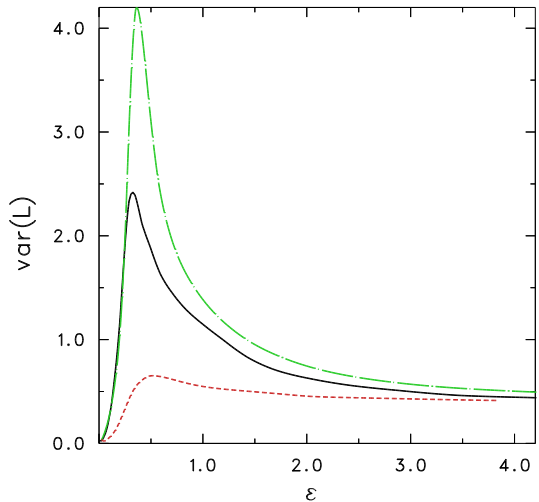}
\caption{The variance of the FPL distribution. Solid black line
  and dashed red line: Z2 liquid for $T=0.7$ and $T=1.2$,
  respectively. Dash-dotted green line, the BLJ liquid.}
\label{fig2}
\end{figure}

\begin{figure} %figure 3
\includegraphics[width=7.9cm]{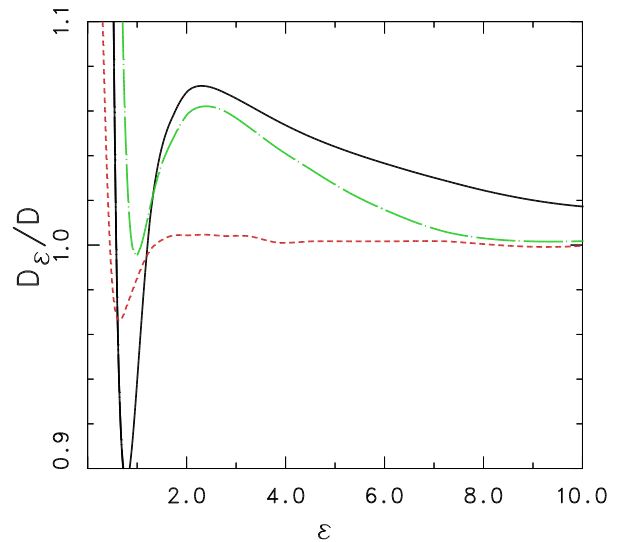}
\caption{ Ratio of the $\varepsilon$-dependent diffusion rate
  (Eq. 3) to the asymptotic Brownian diffusion rate. Solid black
  line and dashed red line: Z2 liquid for $T=0.7$ and $T=1.2$,
  respectively. Dash-dotted green line, the BLJ liquid.}
\label{fig3}
\end{figure}

\begin{figure} %figure 4
\includegraphics[width=7.cm]{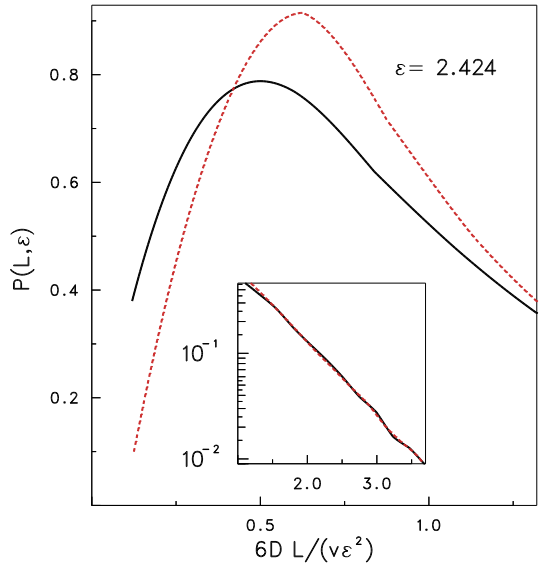}
\caption{ FPL distribution for the particle trajectories in Z2 liquid
  for $\varepsilon=2.424$. $L$ is scaled by $ v \varepsilon^2 / (6D)$,
  the $\varepsilon$-dependent mean first passage trajectory length for
  Brownian diffusion (Eq. 1), $D$ being the asymptotic diffusion rate
  for the respective $T$. Solid line, $T=0.7$; dashed red line,
  $T=1.2$. The inset depicts, in log-linear scale, the large-$L$
  behaviour of the FPL.}
\label{fig4}
\end{figure}

It is worth noting that the observed long-range maximum of
$\alpha(\varepsilon)$ makes the diffusion at shorter scales faster
as compared with the asymptotic Brownian diffusion. To demonstrate
this effect, we introduce the $\varepsilon$-dependent diffusion rate
$D_\varepsilon$ using Eq. 1:
\begin{equation} 
 D_\varepsilon  =   v \varepsilon^2 / (6 L_\varepsilon) 
\end{equation}

Fig. 3 displays the ratio of $D_\varepsilon/D$, where $D$ is the
Brownian diffusion rate derived from the asymptotic behaviour of
the mean-square displacement in the limit of large time (in that
limit, $D_\varepsilon$ and $D$ converge). The general behaviour
of $D_\varepsilon/D$ is related to that of $\alpha(\varepsilon)$,
Fig. 1, by Eq. 2. The long-range maxima of $\alpha(\varepsilon)$
observed for both supercooled liquids, although appear subtle,
obviously result in a measurable reduction of $D_\varepsilon/D$
within the same range of $\varepsilon$, as can be seen in Fig. 3.

To obtain further insight into this effect, we compare in Fig. 4 two
FPL distributions for the Z2 liquid, at $T=0.7$ and at $T=1.2$, in
both cases for $\varepsilon=2.424$. As can be seen in Fig. 3, at
$T=0.7$ that value of $\varepsilon$ represents the diffusion
enhancement regime, whereas for $T=1.2$ the diffusion appears to
attain Brownian behaviour. In order to compare the distributions on
equal ground, we scaled $L$ by $ v \varepsilon^2 / (6D)$, the
$\varepsilon$-dependent mean first passage trajectory length for the
Brownian diffusion (Eq. 1), $D$ being the asymptotic diffusion rate
for the respective $T$. The large-$L$ tails of the distributions
(shown in the inset) are practically indistinguishable, and their
asymptotic behaviour is apparently exponential. The difference in the
first moments of these two distributions arises from a significant
extra contribution from shorter-$L$ trajectories in the case of
$T=0.7$ indicating an excessive presence in the supercooled liquid of
highly mobile particles. This observation is apparently consistent
with both the diffusion enhancement as observed in Fig. 3, and the
excess variance in the FPL distribution within the respective range of
$\varepsilon$, Fig. 2.

A conceptually interesting aspect of these results concerns the
relation between the diffusion and relaxation \cite{Cicerone}. How far
does a particle of a liquid diffuse before the ergodic equilibrium has
been attained? A necessary condition for that is the onset of an
equilibrium (Brownian) diffusion regime in the particle
trajectories. We see in Figs. 1 and 3 that, in a normal liquid state
above $T_A$, Brownian diffusion is reached as soon as a particle
leaves the cage of its neighbours. The appearance of two longer-range
anomalous diffusion regimes in the supercooled liquid state, which we
attribute to the confinement to a metabasin, extends the spatial scale
of the non-equilibrium diffusion process, thereby delaying the relaxation
relative to the diffusion. In this way, the latter anomaly can also be
regarded as an effect of the metabasin topography of the PEL of
fragile liquids.

In summary, we find that two physically and structurally different
supercooled fragile glass-forming liquids demonstrate a long-range
localisation in the particle trajectories. We link this novel effect
to the existence of metabasins in the fragile liquids' PEL
topographies. Based on the results of this study, and our arguments
presented above, the general impact of the PEL metabasins upon the
supercooled liquid dynamics can be described as follows. First,
constraints imposed by the metabasin boundary reduce the number of
accessible degrees of freedom for the intra-metabasin transitions
between the basins of the PEL minima. This induces, within the
respective range of distances, collective dynamics and, consequently,
persistent diffusion with $\alpha(\varepsilon)<2$. Second, on a larger
spatial scale, the system's (time-limited) confinement to a compact
configuration-space domain defined as a metabasin results in a
long-range localisation of the particle trajectories. In this regime,
$\alpha(\varepsilon)>2$. We also show that these anomalies effectively
give an intermediate-range diffusion rate that is faster than the
asymptotic Brownian diffusion. Finally, the existence of the
long-range anomalies in a particle trajectory, delaying its approach
to the equilibrium Brownian diffusion, also delays the liquid's
relaxation relative to diffusion.

The authors thank Prof. D. J. Wales and Dr. V. K. de Souza for a very
useful discussion. The support from the Centre for Parallel Computers
(PDC) is gratefully acknowledged.

\end{document}